\documentclass[prb]{revtex4}
\usepackage{epsf}
\begin{document}
\title{Aspects of Black Hole Physics and Formation of Super-massive Black Holes  from Ultra-light Dark Bosons}
\author{Patrick Das Gupta}
\affiliation{Department of Physics and Astrophysics, University of Delhi, Delhi - 110 007 (India)}
\email{pdasgupta@physics.du.ac.in}
\author{and Fazlu Rahman}
\affiliation{Department of Physics, Cochin University of Science and Technology, Kochi-682 022 (India)}
\vspace{0.1in}

\begin{abstract}

First, we  verify  that the physical parameters estimated for the four directly detected gravitational wave (GW) events involving coalescence of binary black holes (BHs)  indeed uphold the second law of BH thermodynamics, strengthening further the case for  BH physics. Non-spherical gravitational collapse leading to  BH formation may entail very high GW luminosities during the final phase of implosion, reaching non-negligible fraction of  Dyson luminosity $\sim c^5/G $. 

 Most galaxies  harbor  supermassive black holes (SMBHs) in their nuclear regions. Several bright quasars  detected at redshifts  $\gtrsim$ 6 are powered by accreting SMBHs of mass $\gtrsim 10^9 M_\odot$ when the universe was only $\sim 10^9$ yrs old.  We posit that creation of SMBHs   occurs due to collapse  (on dynamical time scales $\sim 10^8$ yrs) of  ultra-light bosonic  dark matter (DM) particles  that have undergone  Bose-Einstein condensation. Furthermore, oscillations in DM Bose-Einstein condensates (BECs) triggered by tidal forces in interacting galaxies  can lead to bursts of  star formation in the galactic nuclei because of frequent collisions of gas clouds due to changing gravitational field. 
 
 We buttress our proposal by first employing simple but tangible physical arguments, and then by  making use of Gross-Pitaevskii equation to study the formation of rotating SMBHs having mass $\gtrsim 10^9 M_\odot$.  We  also make  simple estimates of  GW amplitude as well as luminosity ensuing from the time varying configuration of  BECs, constituted by the ultra-light dark bosons, and remark on the possibility of detecting such GWs using Pulsar Timing Arrays.
\end{abstract}
\maketitle 



\section{Introduction} 
Cosmology, at present, faces three cardinal challenges - explaining the existence of supermassive black holes (SMBHs), dark matter (DM) and dark energy (DE). Almost all galaxies, including    
 our Milky Way, show  evidence of SMBHs  occupying their central regions \cite{pdg-ref-0, pdg-ref-1, pdg-ref-2}. In order to be  bright as well as  display light-curve variability on   time scales as short as few hours, active galactic nuclei (AGNs) like quasars and blazars, require accretion discs around SMBHs that  exploit the deep  gravitational potential of BHs to become hot and luminous \cite{pdg-ref-3, pdg-ref-4, pdg-ref-5}. Numerous SMBHs of mass $\gtrsim 10^9 \ M_\odot $ were likely to have formed  when the universe was barely $\sim 10^9$ yrs old, as  brightest  of the quasars have been detected at high redshifts (z $\gtrsim 6$). \cite{pdg-ref-6, pdg-ref-7, pdg-ref-8, pdg-ref-9, pdg-ref-10, pdg-ref-11}
 J0100+2802, one of the ultra-luminous quasars located at $z= 6.33 $, has  a   SMBH of mass $\sim 1.2 \times 10^{10} \ M_\odot $. \cite{pdg-ref-9} A very recently discovered quasar J1342+0928  located at   $z=7.54$, surpassing the distance of quasar J1120+0641 ($z=7.09$), is estimated to have a SMBH of mass   $\sim 8 \times 10^{8} \ M_\odot $ when the universe was merely $6.9 \times 10^8$ yrs of age \cite{pdg-ref-10, pdg-ref-11}.

Very strong evidence emerging from the  analysis of observed flat rotation curves associated with disc galaxies as well as of  gravitationally bound, rich clusters of galaxies  point to the existence of dark matter (DM)\cite{pdg-ref-12}. Presence of DM  is   inferred also from  gravitational lensing of distant sources by intervening galaxy clusters and from the observed   cosmic microwave background radiation (CMBR) anisotropy \cite{pdg-ref-12, pdg-ref-13}. Cosmological models, like the $\Lambda$CDM,  pertaining to k=0, homogeneous-isotropic universe, with a non-zero cosmological constant and consisting  of weakly interacting massive particles (WIMPs) acting as the cold DM,  are not only consistent with the subtle details of CMBR observations but are also  successful in explaining   cosmic structures  on scales larger than the galactic scales.

 But   predictions of $\Lambda$CDM model at galactic and sub-galactic scales, such as large number of bright satellite galaxies and DM density cusps in the galactic centres    have not met with much success  \cite{pdg-ref-14}. Likewise,  CDM candidates e.g. neutralinos and other SUSY particles  ensuing from  supersymmetric extensions of particle physics Standard Model  have not yet been detected \cite{pdg-ref-15, pdg-ref-16, pdg-ref-17, pdg-ref-18}.

In contrast, if  ultra-light scalar/ pseudo-scalar particles like axion or dynamical four-form  constitute  the DM, there may not arise any tension between the  observed large scale cosmic structures and the  sub-galactic scale features \cite{pdg-ref-14, pdg-ref-19, pdg-ref-20, pdg-ref-21}. Existence of ultra-light, bosonic DM particles may also explain discovery  of SMBHs  at the centers of most galaxies  \cite{pdg-ref-21}. Although growth of seed BHs via matter accretion or direct collapse of galactic halo  are considered to be   standard   scenarios for generating SMBHs, they require either very massive seed BHs  ($\gtrsim 10^3 \ M_\odot $) to have formed by $ z \gtrsim 40 $ or recurring periods of hyper-Eddington accretion rate to amplify the mass to $\gtrsim 10^9 \ M_\odot $  when the age of the universe is only $\sim 10^9$ yrs  \cite {pdg-ref-10, pdg-ref-11, pdg-ref-22, pdg-ref-23, pdg-ref-24, pdg-ref-25, pdg-ref-26, pdg-ref-27, pdg-ref-28, pdg-ref-29, pdg-ref-30, pdg-ref-31}. 

But are there strong  and direct evidence of  BHs? Inference of BH-existence whether  from rapid X-ray variability associated with accretion discs around compact objects like Cygnus X-1 and AGNs,   or from painstaking studies of stellar motions  around  super-massive compact  objects present in numerous galaxy centres, are  indirect. However, recent discovery of gravitational waves (GWs)  by LIGO   detectors have put BHs on very firm footing \cite  {pdg-ref-32,  pdg-ref-33, pdg-ref-34, pdg-ref-35}.

  So,  at the very outset, we make use of the estimated parameters from the directly detected GWs to  further bolster the evidence for BHs. In the section thereafter, we first provide simple arguments to illustrate that self-gravity of a portion of  DM halo, that is in a Bose-Einstein condensate (BEC) phase,   can entail creation of a SMBH. In the subsection that follows, we employ  Gross-Pitaevskii equation to  study the quantum evolution of such  BECs as well as the conditions under which rotating SMBHs can be produced on dynamical time scales. We conclude the section by commenting on the GWs generated during the dynamical changes in such rotating BECs.

\section{Black hole Thermodynamics} 
Black holes (BHs) and gravitational waves (GWs) are two of the signature predictions of general relativity. Although there were plenty of indirect evidence for both in the past,  direct detection have been made possible  by  the LIGO detectors only recently.  Now, for BHs, general relativity also predicts   that no classical process can decrease the area $A$ of a BH event horizon (EH) (i.e. the  second law of BH thermodynamics (SLBHT))  so that \cite {pdg-ref-36},
\begin{equation}
	\frac{dA}{dt}\geq 0 \ ,
\end{equation}
implying that  the EH area $A$ characterizes BH entropy. 

Therefore, according to SLBHT, given two coalescing BHs with  initial EH  area $A_1$ and $A_2$, respectively,  the resulting post-merger compact object must necessarily have an EH area $A \geq A_1 + A_2$, if it settles down to a  standard BH. Recent detection of the four GWs events offer an excellent opportunity to study the significance of SLBHT for these binary BH mergers. In the next  subsection, we check the consistency of  SLBHT by making use of the estimated physical parameters associated with the events - GW150914, GW151226, GW170104 and GW170814. \cite {pdg-ref-32, pdg-ref-33, pdg-ref-34,  pdg-ref-35} The subsection that follows discusses Dyson or Planck luminosity in the context of BH formation.

\subsection{Binary Black hole Mergers from LIGOs} 

The EH area of a  Kerr BH  is given by,
\begin{equation}
    A=8\pi\Bigg(\frac{G}{c^{2}}\Bigg)^{2}M\Bigg(M+\sqrt{M^{2}-\big(Lc/GM\big)^{2}}\Bigg)
\end{equation}
where $M$ and $L$ are the BH's mass and spin angular momentum, respectively. From eq.(2), it is clear  that for a given mass $M$, the EH area is maximum when the BH is of Schwarzschild type (i.e. $L=0$).  In order to make an  extra-stringent test of the SLBHT,  we assume that the two initial BHs are of Schwarszchild kind  so that the initial total EH area is given by,
\begin{equation}
    A_{i}=16\pi\Bigg(\frac{G}{c^{2}}\Bigg)^{2}\Big[({M_{1}^{2}+M_{2}^{2}}\Big]
\end{equation}
\\
BH-BH merging, ensuing from loss of orbital energy energy due to emission of GWs, leads to a bigger rotating BH (because of the orbital angular momenta of the initial BHs) with the final EH area given by,
\begin{equation}
    A_{f}=8\pi\frac{G^{2}}{c^{4}}M_{f}^{2}\Bigg(1+\sqrt{1-\big(Lc/GM_{f}^{2}\big)^{2}}\Bigg)
\end{equation}
From eqs.(3) and (4),  ratio of the final EH area to the initial EH area is simply,
\begin{equation} 
    \frac{A_{f}}{A_{i}}=\frac{M_{f}^{2}\Big[1+\sqrt{1-\big(Lc/GM_{f}^{2}\big)^{2}}\Big]}{2\Big[({M_{1}^{2}+M_{2}^{2}}\Big]} = \frac{M_{f}^{2}\Big[1+\sqrt{1-a_{f}^{2}}\Big]}{2\Big[({M_{1}^{2}+M_{2}^{2}}\Big]}
\end{equation} 
where,
\begin{equation}
a_f \equiv \frac{L}{GM_{f}^{2}/c}\ .
\end{equation}
For the event GW150914, estimated parameters are as follows: $M_{1}=29^{+4.0}_{-4.0} M_{\odot}$, $M_{2}=36^{+5.0}_{-4.0} M_{\odot}$,  $M_{f}=62^{+4.0}_{-4.0} M_{\odot}$ and $a_{f}=0.67^{+0.05}_{-0.07}$.

If we consider  $M_{1}=29 M_{\odot}, M_{2}=36 M_{\odot}, M_{f}=62 M_{\odot}$ and $ a_{f}=0.67$  in eq.(5), we obtain  
$A_{f}/A_{i}=1.57$.

On the other hand, if we take $M_{1}=33 M_{\odot}, M_{2}=41 M_{\odot}, M_{f}=58 M_{\odot}$ and $ a_{f}=0.72$, we find 
    $\big(A_{f}/A_{i}\big)_{min}=1.03$
while with 
$M_{1}=25 M_{\odot},  M_{2}=32 M_{\odot}, M_{f}=66 M_{\odot}$ and $ a_{f}=0.6$, we get 
    $\big(A_{f}/A_{i}\big)_{max}=2.38$
so that  $A_{f}/A_{i}$ may be taken to be $1.57^{+0.81}_{-0.54}$ for GW150914. \cite {pdg-ref-37} In this analysis,  we have made a simplifying assumption that the errors in the estimated parameters, as quoted by Abbott et al. (2016), \cite  {pdg-ref-32} are mutually independent.

Employing the above  method to the three subsequent GW events as well, we arrive at the following table:

\begin{center}
\begin{table}[h]
\begin{tabular}{l | c | c | c | c | p{1 cm}}
\hline

Event  & \textbf{$M_{1}$} & \textbf{$M_{2}$} & \textbf{$M_{f}$} & \textbf{$a_{f}$} & \textbf{$A_{f}/A_{i}$}\\
\hline \hline
{\footnotesize GW150914} & {\footnotesize $ 29^{+4.0}_{-4.0} M_{\odot}$} &  {\footnotesize $36^{+5.0}_{-4.0} M_{\odot}$} & {\footnotesize $62^{+4.0}_{-4.0} M_{\odot}$} &{\footnotesize $0.67^{+0.05}_{-0.07}$} &{\footnotesize $1.57^{+0.81}_{-0.54} $ }\\
\hline
\footnotesize{GW151226} & \footnotesize{$14.2^{+8.3}_{-3.7}$} & \footnotesize{$7.5^{+2.3}_{-2.3}$} & \footnotesize{$20.8^{+6.1}_{-1.7}$} & \footnotesize{$0.74^{+0.06}_{-0.06}$} & \footnotesize{ $1.40^{+3.16}_{-0.92}$}\\
\hline

{\footnotesize GW170104} & {\footnotesize$31.2^{+8.4}_{-6.0} M_{\odot}$} &\footnotesize{ $19.4^{+5.3}_{-5.9} M_{\odot}$} & \footnotesize{$48.7^{+5.7}_{-4.6} M_{\odot}$} &\footnotesize{ $0.64^{+0.09}_{-0.20}$} & \footnotesize{ $1.55^{+1.88}_{-0.8}$}\\
\hline
\footnotesize{GW170814} & \footnotesize{$30.5^{+5.7}_{-3.0} M_{\odot}$} & \footnotesize{$25.3^{+2.8}_{-4.2} M_{\odot}$} &\footnotesize{ $53.2^{+3.2}_{-2.5} M_{\odot}$} &\footnotesize{ $0.70^{+0.07}_{-0.05}$}&\footnotesize{ $1.54^{+0.78}_{-0.54}$}\\
\hline
\end{tabular}
\end{table}
\end{center}
It is evident from the table that for all the  four events, average $A_f/A_i> 1$, which is consistent with the SLBHT. However, minimum values of $A_f/A_i$ are less than unity for GW151226 and GW170104,  suggesting  that the initial BHs for these events were not of Schwarzschild kind if $M_1$,  $M_2$ and $a_f$  actually  corresponded to the largest allowed values while $M_f$ to the minimum.

\subsection{Planck Scales, Dyson Luminosity  and Hawking Radiation}  
The gravitational wave luminosity, due to a source with slow internal motion, is given by \cite {pdg-ref-38},
\begin{equation}
L_{GW}=\frac {G} {2 c^5}  \bigg < \dddot{\not{I}}_{jk}\dddot{\not{I}}^{jk} \bigg > 
\end{equation}
where, 
$$\not{I}^{ij}(t) \equiv I^{ij}(t) - \frac {1} {3} \delta^{ij} I^k_k (t)$$
is the reduced mass quadrupole moment, with  mass quadrupole moment $I^{ij} (t)$ defined by,
$$I^{ij} (t) \equiv \int {\rho(t, \vec r) x^i x^j d^3r} \ \ ,$$
 $\rho(t, \vec r)$ being the  mass density of the  source.
 
It is  interesting to note that  physical dimensions of both $\frac { c^5}  {G}=3.6 \times 10^{59} \mbox{\ erg \ s}^{-1}$ and $\dddot{\not{I}}_{jk}$ are identical.  Furthermore, if one considers Planck energy, $E_{Pl} \equiv m_{Pl} c^2 \equiv \sqrt{c^5 \hbar / G}$ (where, $m_{Pl} \equiv \sqrt{\hbar c/G}$ is the Planck mass) and the Planck time, $t_{Pl} \equiv \sqrt{\hbar G/ c^5}$ to define  Planck luminosity  (or, equivalently, Dyson luminosity) \cite {pdg-ref-38, pdg-ref-39, pdg-ref-40, pdg-ref-41},
\begin{equation}
L_{Pl}\equiv \frac {E_{Pl}}{t_{Pl}}=\frac{c^5} {G} 
\end{equation}
then one concludes that, around the time of big bang, quantum fluctuations could have generated GWs with luminosity $\sim c^5/G$ and that, since the quantum imprint $\hbar $ is missing from the RHS in eq.(8), it is preferable that  $\sim c^5/G$ be termed as Dyson luminosity (as Dyson was the first to discuss it in detail \cite {pdg-ref-39, pdg-ref-41}). 
 
Suppose the mass distribution in a GW source varies asymmetrically over a typical time scale $ \tau \sim  2 \pi/\omega=1/f$, $f$ being the characteristic GW frequency, while the macroscopic non-spherical internal kinetic energy associated with the source is   $E_{nonsph}$,  then one can make simple estimate of GW luminosity  using eq.(7),
\begin{equation} 
L_{GW} \sim \frac {2 G} { c^5} \omega^2 E^2_{nonsph} \approx 2 \times 10^{50} \ \bigg (\frac {E_{nonsph}} {10^{51}\ \mbox{erg\ s}^{-1}} \bigg )^2 \bigg ( \frac {f} {1\ \mbox{kHz}} \bigg )^2 \ \ \mbox{erg\ s}^{-1}\ \ .
\end{equation}
One   may consider the case of gravitational collapse of a compact cosmic object like a supra-massive neutron star or an over-dense pocket in the early universe of mass $M$ and initial size $R=\alpha_1 R_s$, where $ R_s\equiv 2 G M/c^2$ is the Schwarzschild radius and $\alpha_1 \gtrsim 1$. Non-spherical collapse of such a system to a BH  with  asymmetric kinetic energy $E_{nonsph}= \alpha_2 M c^2$ ($\alpha_2 \lesssim 1$) occurring on a dynamical time scale $\sim \sqrt{R^3/GM} \sim \tau_{dyn}$  would, according to eq.(9), lead to a GW luminosity,
\begin{equation} 
L_{GW}\sim \frac{\alpha^2_2} {4 \alpha^2_1}\ \frac { c^5} {G}= 9 \times 10^{58} \ \frac{\alpha^2_2} { \alpha^2_1}\  \mbox{erg \ s}^{-1}
\end{equation}
while  $\dddot{\not{I}}_{jk} \sim   E_{nonsph}/\tau_{dyn}\sim \frac{\alpha_2} {2 \sqrt{2} \alpha_1}\ \frac { c^5} {G}$  is essentially  of the same order as $L_{GW}$ provided $\alpha_1$ and $\alpha_2$ are of order unity.

From eq.(10) it ensues that collision of bubble walls or rapid collapse of false vacuum pockets in the early universe \cite {pdg-ref-42, pdg-ref-43} could generate GW luminosity that is non-negligible fraction of   $c^5/G$. Similarly, supra-massive neutron stars or magnetars (that are plausible  progenitors of fast radio bursts/long gamma ray bursts  \cite {pdg-ref-44}) collapsing to form BHs, as they lose the centrifugal support due to magnetic braking, could also lead to $L_{GW} \lesssim c^5/G$.  However, in all such situations, the total energy carried away by GWs is limited by $E_{nonsph}=\alpha_2 M c^2$, even though  GW luminosity may  approach  the Dyson luminosity from below.

Special relativity and causality arguments constrain the initial source size to $R \lesssim c \tau_{dyn}$. When this constraint is applied to eq.(9) we obtain, 
\begin{equation} 
L_{GW} \sim \frac { G} { c^5} \frac {E^2_{nonsph}} {\tau^2_{dyn}} \lesssim \frac { G c M^2} {R^2} \ .  
\end{equation}
Therefore, one arrives at an upper limit for $L_{GW}$ by substituting the smallest possible size $R_{min} \sim G M/c^2$ in eq.(11),
\begin{equation}
L_{GW} < \frac { G c M^2} {R^2_{min}}=  \frac{c^5} {G}\ , 
\end{equation}
which indicates that Dyson luminosity may represent an upper limit for the  GW luminosity \cite {pdg-ref-38, pdg-ref-39, pdg-ref-40, pdg-ref-41}. Indeed, the peak luminosity in the case of GW150914 does respect this constraint \cite {pdg-ref-45}. 

Eq.(12), combined  with $E_{nonsph}=\alpha_2 M c^2$ and $L_{GW} \sim E_{nonsph}/ \tau_{dyn}$, entails a lower limit for the time scale  $\tau_{dyn}$ (or equivalently, an upper limit on the characteristic frequency $f$) on which GW source matter gets redistributed,
\begin{equation}
 \tau_{dyn} \geq \frac {\alpha_2 G M} {c^3}=1.5 \times 10^{-5} \ \alpha_2 \ \bigg (\frac {M} {3 \ M_\odot} \bigg )\ \mbox{s} 
\Rightarrow f \leq \frac {c^3}{\alpha_2 G M}\ \cong  67 \ \alpha^{-1}_2 \ \bigg (\frac {M} {3 \ M_\odot} \bigg )^{-1} \ \mbox{kHz} \ \ .
\end{equation} 
For GW150914, if one takes the mass scale to be $\sim 30-40 \ M_\odot$, one does find that the observed maximum frequency $\sim 150$ Hz \cite{pdg-ref-45} is less than the upper limit deduced from eq.(13).

Luminosity associated with  the  evaporation of a BH of mass $M$,  as it  radiates away energy with  flux $F_H=\sigma T^4_H$, where $T_H = c^3 \hbar /8 \pi G M k_B$ is the Hawking temperature, is given by \cite{pdg-ref-46},  
\begin{equation} 
L_H \approx F_H\times 4 \pi R^2_s=\frac {1} {15360 \pi} \bigg (\frac {m_{Pl}} {M} \bigg )^2 \frac{c^5}{G} 
\end{equation} 
This implies that even for a Planck mass primordial BH, the Hawking  luminosity  (eq.(14)) is four orders of magnitude smaller than the Dyson bound, which is understandable since Hawking radiation is manifestly driven by quantum effects while Dyson luminosity is devoid of $\hbar $.

\section{Bose-Einstein Condensation of Ultra-light Dark Bosons and Formation of Massive Black Holes} 
A very recent study, discussing SMBHs associated with three  most distant quasars, J0100+2802 ($z=6.33$),  J1120+0641 ($z=7.09$) and J1342+0928 ($z=7.54$), has arrived at the conclusion that scenarios in which   BHs  grow by accreting matter to become SMBHs with mass $\gtrsim 10^8 \ M_\odot $ as early as $z \gtrsim 7$,  need  presence of  very heavy seed BHs (mass $\gtrsim 10^3 \ M_\odot$)  at $z \gtrsim 40$. \cite {pdg-ref-11}  Is there an alternate scenario that can produce SMBHs early on by circumventing such stringent requirements?  In this section, we describe a model  that invokes ultra-light, bosonic DM particles to generate SMBHs \cite{pdg-ref-21}. 
    
As discussed in section I, existence of DM is inferred from the observed rotation curves, virialized galaxy clusters, gravitational lensing by galaxies/clusters and closely interacting systems of galaxies \cite{pdg-ref-12, pdg-ref-13}. We explore the consequence of Bose-Einstein condensation of ultra-light scalar/ pseudo-scalar particles like axion or dynamical four-form  making up the DM  \cite{pdg-ref-14, pdg-ref-19,  pdg-ref-20, pdg-ref-21}. A dynamical four-form can not only entail DE and ultra-light DM particles but may also lead to magnetogenesis  through Chern-Simons extension of electrodynamics \cite {pdg-ref-47}. 

Here, we propose that rotating SMBHs  ensue from the gravitational collapse of BEC made of ultra-light dark bosons  having  non-zero angular momenta. To obtain a physical insight into our proposal, we first employ basic physics to understand the mechanism underlying the theory before going into    studying   the problem  that involves solving a non-linear Gross-Pitaevskii equation.
\subsection{Dark matter condensates,  uncertainty principle and the central region of galaxies}
Since the universe cools down as it expands, it is easy to see that  the  thermal de Broglie wavelength of identical bosons, distributed uniformly on cosmological scales and  having rest mass $\lesssim 1$ eV, is  larger than the mean separation  between them, at all epochs. Hence, for such bosons, the critical temperature to undergo Bose-Einstein condensation is always greater than the temperature of the universe \cite{pdg-ref-48, pdg-ref-49}.

When such light, non-relativistic and weakly interacting bosons of mass $m$ constitute DM halo of size $R_h$, a fraction of them with very low momenta $p$ can form a condensate  provided,
\begin{equation}
\lambda_{DB} \sim \frac {h} {p} \gtrsim \bigg (\frac {3 N} {4 \pi R^3_h} \bigg )^{-1/3}= R_h \ \bigg (\frac {3 M} {4 \pi m } \bigg )^{-1/3} \ ,
\end{equation}
where $N$ and $M$ are the number and the total mass of dark bosons making up the BEC. If the DM halo develops an angular momentum  due to tidal torques resulting from  encounters with other galaxies \cite {pdg-ref-50, pdg-ref-51}, then the energy of a typical boson, that is a part of the rotating BEC and has an orbital angular momentum $l$,  is simply,
\begin{equation}
E \sim \frac {p^2} {2m} + \frac {l^2} {2m R^2_h} - \frac {G M m} {R_h} < 0
\end{equation}
so that,
\begin{equation}
p^2 < \frac {2 G M m^2} {R_h} - \frac {n^2 \hbar^2} {R^2_h}
\end{equation}
after employing  $l\sim n \hbar, \ n=1,2,...$, as ordained by quantum theory, in eq.(16). The total angular momentum of the BEC is, of course,
\begin{equation}
L \sim n N \hbar = n  \hbar \bigg (\frac {M} {m} \bigg )
\end{equation} 

Heisenberg's uncertainty principle demands that,
\begin{equation}
\Delta p \sim p \gtrsim \frac {\hbar} {2 R_h}.
\end{equation}
According to eqs.(15) and (19), the size of the gravitationally bound BEC satisfies the inequality,
\begin{equation}
\frac {h} {4 \pi p} \lesssim R_h \lesssim \frac {h} {p} \bigg (\frac {3 N} {4 \pi} \bigg ) \ .
\end{equation}
The above condition, in the case of a BEC, is self-consistent as $N \gg 1$.

From minimum energy consideration, eq.(19) entails $p \sim \frac {\hbar} {2 R_h}$ be substituted in eq.(16) so that,
\begin{equation}
E \sim \frac {\hbar^2} {8 m R^2_h} + \frac {n^2 \hbar^2} {2m R^2_h} - \frac {G M m} {R_h} < 0
\end{equation}
implies,
\begin{equation}
R_h \gtrsim \bigg (n^2+ \frac {1}{4} \bigg ) \frac {\hbar^2} {2 G M m^2} = 0.5 \bigg ( n^2+\frac {1}{4}  \bigg )   \bigg ( \frac {m^2_{Pl}}{m\ M} \bigg )  \bigg ( \frac {\hbar} {m c} \bigg ) = 4.3 \bigg ( n^2+\frac {1}{4}  \bigg )   \bigg ( \frac {10^7 M_\odot}{M} \bigg ) \bigg ( \frac {10^{-22} \ \mbox{eV}}{m} \bigg )^2 \ \mbox{kpc}
\end{equation}
For a fixed angular momentum, $L= n N \hbar$, one can obtain an estimate of the BEC size by further minimizing $E$ with respect to $R_h$ by setting,
\begin{equation}
\frac{\partial E}{\partial R_h}=\frac{G M m}{R^3_h} \bigg [R_h - \bigg ( n^2+\frac {1}{4}  \bigg ) \frac {\hbar^2}{G M m^2} \bigg ]=0
\end{equation}
so that the size $R_{h0}$ that leads to minimum energy configuration for the BEC is given by,
\begin{equation}
R_{h0}=\bigg ( n^2+\frac {1}{4}  \bigg ) \frac {\hbar^2}{G M m^2} \cong 86 \ \bigg ( n^2+\frac {1}{4}  \bigg )   \bigg ( \frac {10^9 M_\odot}{M} \bigg ) \bigg ( \frac {10^{-22} \ \mbox{eV}}{m} \bigg )^2 \ \mbox{pc}
\end{equation}
corresponding to a single boson energy,
\begin{equation}
E_{min}= - \frac {G M m} {2 R_{h0}} = - 0.5 \ \bigg ( \frac {m c^2} {n^2 + 1/4} \bigg )  \bigg ( \frac {m^2_{Pl}}{m\ M} \bigg )^{-2} \ .
\end{equation}
Eq.(24) tells us that larger the angular momentum and smaller the total mass of the BEC, bigger is the latter's size. 
However, for a fixed angular momentum,  with increasing BEC mass $M$ not only its size decreases the corresponding Schwarzschild scale $R_s$ increases, making the possibility of irreversible gravitational collapse inevitable.

Any gravitational perturbation (e.g. galaxy-galaxy interaction) would tend to make the size of the condensate  oscillate about $R_{h0}$. A characteristic normal mode  frequency $\omega $ can easily be obtained within the framework of small oscillations. Beginning with,
\begin{equation}
R_h(t)= R_{h0} + x(t)
\end{equation}
where $\vert x(t) \vert \ll R_{h0}$ is the amplitude of oscillation. Substitution of eq.(26) in eq.(21) leads to,
\begin{equation}
E = E_{min} + \frac {x^2} {R^3_{h0}} \bigg [\frac {3 \hbar^2 ({n^2 + 1/4})} {2 m R_{h0}} - G M m \bigg ]=  E_{min} + \frac {1} {2} \frac{ G M m} {R^3_{h0}} x^2
\end{equation}
From eq.(27), it is evident that the small oscillations about $R_{h0}$ are executed with a characteristic frequency,
\begin{equation}
\omega=\sqrt{\frac{ G M } {R^3_{h0}}}
\end{equation}
corresponding to a time period $\tau=\frac{2 \pi} {\omega}$,
\begin{equation}
\tau=2 \pi ({n^2 + 1/4})^{3/2}\bigg ( \frac {1}{G M} \bigg )^2  \bigg ( \frac {\hbar} {m } \bigg )^3 \cong 2 \times 10^6 \ ({n^2 + 1/4})^{3/2} \bigg (\frac {m} {10^{-22}\ \mbox{eV}} \bigg )^{-3}\bigg (\frac {M} {10^9\ M_\odot} \bigg )^{-2}\ \mbox{yrs}
\end{equation}
In general, the gravitational potential due to a rotating BEC would not be spherically symmetric like the one given by eq.(16). However, the time scale of small oscillations is not going to be very different from the expression given by eq.(29) if the spinning BEC is not highly deformed.
In that case,  undulations in the dark matter BEC triggered by galactic encounters can have interesting  astrophysical implications since  eqs.(24) and (29) suggest that the gravitational potential can vary on time scales of $\sim 10^6 $ yrs in a region of size $\lesssim 100$ pc.

It is well known that star burst galaxies as well as host galaxies of quasars  and other AGNs  exhibit  evidence of not only  
very high star formation rates (SFRs) compared to those in normal galaxies but also of close encounters with other galaxies \cite{pdg-ref-52, pdg-ref-53, pdg-ref-54, pdg-ref-55}. Very rapid creation of stars out of available gas  on a time scale  of $\sim 10^7 $ yrs  is often confined to the   nuclear regions of size $\lesssim 100$ pc in star burst galaxies \cite{pdg-ref-52}.  Furthermore, the host galaxy of a recently discovered  distant quasar J1342+0928 ($z = 7.54$) displays a high SFR  of about  $85-345\ M_\odot\ \mbox{yr}^{-1}  $ along with an estimated mass of $(0.6-4.3)\times 10^8 \ M_\odot$  in dust\cite{pdg-ref-10}. 

 Now, energy of individual gas clouds or stars is not  conserved when the gravitational potential varies with time, as discussed in  seminal papers by Lynden-Bell in the context of violent relaxation of stellar systems \cite{pdg-ref-56, pdg-ref-57}. The ensuing crossing of orbits  due to time varying potential, because of the changing BEC size in our scenario,  entails   frequent collisions  of  gas clouds \cite{pdg-ref-58}.  Collisions would  cause sudden  compression of gas clouds, driving shock waves through them,  leading to gravitational instabilities and   formation of stars thereby \cite{pdg-ref-59}. 
 
Such shocks could enhance a top heavy SFR on time scales $\sim 10^6 -  10^7 $ yrs, and  plausibly   explain the observed rapid rates of star formation  in star burst galaxies and presence of large amount of dust when the universe is only $\sim 10^8-  10^9$ yrs old.   

Eventually the excited dark matter BEC would lose its energy to stars and gas clouds because of the time varying gravitational potential and settle down to a lower energy configuration like what was obtained in eqs.(21)-(25). The BEC will implode to form a BH  if its size $R_{h0}$ is less than the Kerr EH radius  given by,
\begin{equation}
R_{BH}=\frac {R_s} {2} + \sqrt{\bigg (\frac {R_s} {2} \bigg )^2 - \bigg (\frac 
{L}{Mc} \bigg )^2} \ .
\end{equation}
By making use of the condition $R_{h0} \lesssim R_{BH}$ along with eqs.(18), (24) and (30), we may express the criteria for the BH formation to be,
\begin{equation}
 \bigg (n^2+ \frac {1}{4} \bigg ) \frac {\hbar^2} { G M m^2}  \lesssim \frac { G M} {c^2}  \bigg [1+ \sqrt{1- \bigg (\frac{n \hbar c} {G M m} \bigg )^2} \ \bigg ]
\end{equation}
so as to obtain the inequality,
\begin{equation}
\bigg (n^2+ \frac {1}{4} \bigg ) \bigg (\frac{ m^2_{Pl}} {M \ m} \bigg )^2 \lesssim 1+ \sqrt{1- \bigg (\frac{n \ m^2_{Pl}} {M \ m} \bigg )^2} \ .
\end{equation}
In order to avoid appearance of imaginary numbers in the RHS above, eq.(32) demands that,
\begin{equation}
m \ M \geq n \ m^2_{Pl}
\end{equation}
From the inequality in eq.(32), it follows that,
\begin{equation}
 m \ M \gtrsim   \frac {n^2 + 1/4}{\sqrt{n^2 + 1/2}}  \  m^2_{Pl}  
\end{equation}
which automatically subsumes the reality criteria ensuing from eq.(33). For n=1, the above result implies,
\begin{equation}
 m \ M \gtrsim  1.02  \  m^2_{Pl}  
\end{equation}

In what follows, we will derive a more accurate constraint by studying the evolution of DM in the BEC phase using the framework of Gross-Pitaevskii equation.

\subsection{Gross-Pitaevskii Equation and Ultra-light Dark Matter Particles}
If the  galactic halos are made of ultra-light dark bosons (rest mass $\lesssim $ 1 eV), a very large fraction of  them can be in  BEC states \cite {pdg-ref-14, pdg-ref-60}.  Typical speed of such particles in a  DM halo is  $\lesssim $ 100 km/s, so that a non-relativistic study of the quantum problem is adequate for all practical purposes. 

In the mean field approximation, dynamics of the  BEC is governed by the time evolution of the condensate wavefunction.  Then,  evolution of the condensate wavefunction  $\psi (\vec{r},t)$   can be described  by the following Gross-Pitaevskii equation (GPE),
\begin{equation}
i \hbar {{\partial \psi}\over{\partial t}}  = \bigg [- \frac {\hbar^2}{2m} \nabla ^2    + V_{ext} + N \int {V(\vec{r} - \vec{u}) \vert \psi (\vec{u},t)\vert ^2 d^3u} \bigg ] \psi (\vec{r},t) 
\end{equation}
where,
\begin{equation}
 V(\vec{r} - \vec{u})= \frac {4 \pi \hbar ^2  a} {m} \delta ^3 (\vec{r} - \vec{u}) + V_g (\vert \vec{r} - \vec{u} \vert)
\end{equation} 
so that,
$$i \hbar {{\partial \psi}\over{\partial t}} = \bigg [- \frac {\hbar^2}{2m} \nabla ^2    + V_{ext} + N g  \vert \psi (\vec{r},t)\vert ^2 + $$
\begin{equation}
  + N \int {V_g (\vert \vec{r} - \vec{u} \vert) \vert \psi (\vec{u},t)\vert ^2 d^3u} \bigg ] \psi (\vec{r},t) 
\end{equation}  
where $m$ is the dark  boson rest mass, $g \equiv \frac {4 \pi \hbar ^2  a} {m}$ characterizes a short range contact interaction between the bosons and 
$V_{ext} (r)$ is the gravitational energy due to a compact remnant of mass $M_0$ at the centre  (plausibly originating from a population III star)  given by,
\begin{equation}
 V_{ext} (r) =  - \frac {G M_0 m}  {r}\ \ \ \ .
\end{equation} 

The GPE of eq.(38) can be derived by extremizing the following  action,
\begin{equation} 
  S = \int {dt \int {d^3r \ \mathcal{L}}} 
\end{equation}  
 where,
$$\mathcal{L}= \frac {i \hbar}{2} \bigg \lbrace  \psi {{\partial \psi^*}\over{\partial t}} - \psi^* {{\partial \psi}\over{\partial t}}\bigg \rbrace + \frac {\hbar^2}{2m}   \nabla \psi^*. \nabla \psi + V_{ext}\ \vert \psi \vert ^2 + $$ 
\begin{equation} 
\ \ \ \ \ \ \ + \frac {g N} {2} \vert \psi \vert ^4 + \frac {N} {2} \vert \psi \vert ^2 \int {V_g (\vert \vec{r} - \vec{u} \vert) \vert \psi (\vec{u},t)\vert ^2 d^3u}  \ \ \ . 
\end{equation}

Of course, the mutual gravitational interaction between any pair of ultra-light dark bosons separated by a distance $r$ is given by,
\begin{equation}
V_g (r) = - \frac {G m^2}  {r} \ \ \ \ \ . 
\end{equation}

Substituting eqs.(37), (39) and (41) in the  action given by eq.(40) results in,
$$ S = \int {dt \int {d^3r \ \mathcal{L}}} = \int {dt \int {d^3r \  \psi^* \bigg \lbrace  - {i \hbar}  {{\partial \psi}\over{\partial t}} - \frac {\hbar^2}{2m} \nabla^2 \psi + V_{ext}\ \psi}} 
 + $$ 
\begin{equation} 
 \ \ \ \ \ \ \ + \frac {g N} {2} \vert \psi \vert ^2  + \frac {N} {2}  \int {V_g (\vert \vec{r} - \vec{u} \vert) \vert \psi (\vec{u},t)\vert ^2 d^3u} \bigg \rbrace \psi 
\end{equation}

From here onwards, we will set $g=0$, assuming that the contact interaction between the dark bosons is extremely weak. Now, to obtain an approximate solution of eq.(38) we make use of the time dependent variational method by employing a trial wavefunction (normalized to unity) \cite {pdg-ref-61, pdg-ref-62, pdg-ref-63},
\begin{equation}
\psi(\vec{r},t)= A(t)\ r \ \exp{(-r/\sigma(t))} \exp {(- i B(t)\ r)}\ Y_{l m} (\theta, \phi)
\end{equation}
The normalization condition entails $A(t)$ and $\sigma (t)$ to be related by,
\begin{equation}
\vert A(t) \vert^2= \frac {4} {3}  (\sigma(t))^{-5} \Rightarrow A(t)= \frac {2} {\sqrt{3}}(\sigma(t))^{-5/2} 
\end{equation}
so that  the Lagrangian density of eq.(41) leads to a Lagrangian,
\begin{equation}
L= \int {d^3r \ \mathcal{L}} =  -\ \frac {5} {2} \hbar \sigma \dot{B}  + \frac{\hbar^2}{2m} B^2 + \frac{\hbar^2}{2 m \sigma^2} + L_{int} \ - \ \frac {G M_0 m}  {2\sigma}
\end{equation}
where the self-gravity term $L_{int}$ is given by,
\begin{equation}
L_{int} \equiv \frac {N} {2} \int {d^3r \vert \psi (\vec{r},t) \vert ^2 \int {V_g (\vert \vec{r} - \vec{u} \vert) \vert \psi (\vec{u},t)\vert ^2 d^3u}} \ .
\end{equation}

In the scenario envisaged in this article,  a BEC corresponding to a very  large number, $N$, of dark bosons with momentum $\ll m c$, spread initially  over a  galactic scale $\sim $ 20 - 30 kpc, evolves quantum mechanically according to the GPE of eq.(38) with $g=0$.
  
If we choose $l=1$ and $m=1$ so that,
\begin{equation}
\psi(\vec{r},t)= A(t)\ r \ \exp{(-r/\sigma(t))} \exp {(- i B(t)\ r)}\ Y_{1 1} (\theta, \phi)
\end{equation}
then the self-gravity term of eq.(47) is given by,
\begin{equation}
L_{int}  = -\ \frac {0.37 N G m^2}  {2\sigma} \ \ \ \ .
\end{equation} 
The general case, involving arbitrary  values of $l$ and $m$, will be described in a separate paper \cite {pdg-ref-64}.
Using eqs.(48) and (49) in the Lagrangian given by eq.(46), one derives the following 
Euler-Lagrange equations,  
\begin{equation}
B(t)= - \ \frac{5 m\ \dot{\sigma}}{2\hbar}
\end{equation} 
and,
\begin{equation}
\frac {5} {2} \hbar \dot{B} + \frac {\hbar^2}{ m \sigma^3}  - \frac {G [0.37 Nm +  M_0]m} {2 \sigma^2}=0
\end{equation}
so that the above two equations can be combined to yield,
\begin{equation}
m \ddot{\sigma} = - \frac{dV_{eff}} {d\sigma}
\end{equation}
where,
\begin{equation}
V_{eff}\equiv \frac {2} {25} \bigg [\frac {\hbar^2}{ m \sigma^2} - \frac {G(0.37 Nm + M_0)m}{\sigma}\bigg ]
\end{equation}


Eq.(52) can be trivially integrated to obtain,
\begin{equation}
\frac {1}{2} m {\dot {\sigma}}^2 + V_{eff} = \mbox {Constant} \equiv K_0   \Rightarrow \dot{\sigma}=\pm \frac {2} {5} \sqrt {\frac {G \bar{M}} {\sigma} - \frac {\hbar^2}{m^2 \sigma^2}+ \frac{25} {2} K_0}
\end{equation}
where,
$$\bar {M} \equiv 0.37 Nm + M_0 $$
The time evolution of the trial wavefunction is completely determined by specifying the initial data $\sigma (t_i)$  and $\dot {\sigma} (t_i)$ at an initial time $t_i$. 


If  at time $t_i$, the condensate wavefunction is spread over a galactic scale and is contracting   ever so slowly, implying  $2 \sigma _i \equiv 2 \sigma (t_i) \approx $ 25 kpc and $\dot {\sigma} (t_i) = \ -\ \epsilon $ where $\epsilon \approx 0$ then,
$$K_0 \approx 0  \ \ ,$$ 
so that  one can easily integrate eq.(54) to obtain,

$$ t - t_i= \frac {5} {3} \sqrt{\frac { \sigma^3_i} {G \bar {M}}} \bigg [(1-\frac {\hbar^2}{G \bar{M} m^2 \sigma_i} )^{3/2} - (\frac {\sigma(t)} {\sigma_i} -\frac {\hbar^2}{G \bar{M} m^2 \sigma_i} )^{3/2} \bigg ] - $$
\begin{equation}
- \frac {2\hbar^2}{G \bar{M} m^2 \sigma_i}\bigg [(1-\frac {\hbar^2}{G \bar{M} m^2 \sigma_i} )^{1/2} - (\frac {\sigma(t)} {\sigma_i} -\frac {\hbar^2}{G \bar{M} m^2 \sigma_i} )^{1/2} \bigg ]
\end{equation}

The turning point occurs at $\sigma_{min}$ corresponding to $\dot {\sigma}$ =0 so that,
\begin{equation}
\sigma_{min}= \frac {\hbar^2} {G \bar{M} m^2}
\end{equation}
In general, after reaching the turning point, $\sigma (t) $ starts increasing again following a bounce. However, in those situations in which the contracting BEC size $\approx 2\sigma (t)$ becomes comparable to the associated event horizon radius (eq.(30)),  general relativistic effects start taking over and as a result there is no bounce, instead the size keeps shrinking.  In order to find out the conditions under which the BEC  implodes into a BH, we adopt a heuristic approach by first considering 
the total mass enclosed within a sphere of radius $2 \sigma (t)$ at time $t$,
$$M_0 + M_{\psi}(<2\sigma (t), t)= M_0 + N m \int^{2 \sigma (t)}  _0 \int^{2\pi}_0 \int^{\pi}_0  {\vert \psi(r,t)\vert ^2 d^3r}$$
\begin{equation}
= M_0 + \frac{ 4  N m} {3 (\sigma(t))^5} \int^{2 \sigma (t)} _0{r^4 \exp{(- 2 r/\sigma(t))} dr} \cong M_0 + 0.37 N m  = \bar {M}
\end{equation}
Then, the corresponding angular momentum of the condensate is given by,
\begin{equation}
\bar{L} = N \hbar \int^{2 \sigma (t)} _0 \int^{2\pi}_0 \int^{\pi}_0  {\vert \psi(r,t)\vert ^2 d^3r}= 0.37 N \hbar \ \ .
\end{equation}
Hence, when  $\bar {M}$ and  $\bar {L}$ of eqs.(57) and (58) are substituted as  effective mass and angular momentum, respectively, in the EH radius given by eq.(30), we get,
\begin{equation}
R_{BH}= \frac {G\bar {M}} {c^2} \bigg [ 1 + \sqrt{1 - \frac {m^4_{pl}} {\bar {M}^2 m^2}(1- M_0/\bar {M})^2}\bigg ]
\end{equation}


The dark boson condensate will certainly collapse to form a BH if,
\begin{equation}
2 \sigma_{min}= \frac {2\hbar^2} {G \bar{M} m^2} < R_{BH}
\end{equation}

Therefore, from eq.(60), we have the following criteria for the formation of a BH of mass $\bar{M}$:
\begin{equation}
 m  \bar {M}  > \frac {m^2_{Pl}} {\sqrt{1 - \frac {1} {4} (1- M_0/\bar {M})^2}} \cong 1.15  m^2_{Pl} 
\end{equation}
It is interesting to note that the condition given by eq.(35), although derived using simple physical arguments  in the previous subsection, is not much different from the above inequality.

So, eqs.(55) and (61) imply that the dark boson mass must satisfy,
\begin{equation}
m > 1.54 \times 10^{-20} \ \bigg (\frac{\bar{M}}{10^{10} \ M_\odot} \bigg ) \ \mbox{eV}
\end{equation}
in order that  SMBHs heavier than billion solar masses are formed on time scales,
\begin{equation}
\tau_{dyn}= t-t_i \cong  \frac {5} {3} \sqrt{\frac { \sigma^3_i} {G \bar {M}}} \approx 10^8 \ - \ 10^9 \ \mbox{yrs} \ \ ,
\end{equation}
assuming  an initial BEC size $2 \sigma _i  \approx $ 20-30 kpc.

In this scenario, lighter ($m \sim 10^{-23} \ \mbox{eV}$) DM particles can  even lead to formation of SMBHs of mass $> 10^{12} \ M_\odot$. If the fraction of DM halo that undergoes Bose-Einstein condensation does not vary appreciably from galaxy to galaxy, then the above process can naturally explain the observed  correlation between the mass of the SMBH and the halo mass  \cite {pdg-ref-65}. 




It is also possible to make simple estimates of  low frequency  gravitational radiation that is emitted as  the size of  the condensate changes with time. Assuming an asymmetric variation in size,  energy carried away by gravitational radiation is, 
\begin{equation}
E_{GW} \approx \epsilon \ \frac {G {\bar{M}}^2} {\sigma_{min}} =  \epsilon \ 2.4 \times 10^{64} \  \ \bigg (\frac{\bar{M}}{10^{10} \ M_\odot} \bigg )^3 \ \bigg (\frac{m}{1.54 \times 10^{-20} \ \mbox{eV}} \bigg )^2 \mbox{erg}
\end{equation}
over a time scale given by eq.(63),
\begin{equation}
\tau_{dyn} =   9.4 \times 10^8  \ \bigg (\frac{\bar{M}}{10^{10} \ M_\odot} \bigg )^{-1/2} \ \bigg (\frac{\sigma_i}{25 \ \mbox{kpc}} \bigg )^{3/2} \  \mbox{yrs}
\end{equation}
where the parameter $\epsilon $ characterizes the asymmetric changing size.

Hence, the  gravitational wave luminosity can be estimated to be,
\begin{equation}
L_{GW} \sim \frac{E_{GW}}{\tau_{dyn}}
\approx \epsilon \ 10^{48}  \bigg (\frac{\bar{M}}{10^{10} \ M_\odot} \bigg )^{7/2} \ \bigg (\frac{m}{1.54 \times 10^{-20} \ \mbox{eV}} \bigg )^2 \ \bigg (\frac{\sigma_i}{25 \ \mbox{kpc}} \bigg )^{-3/2}\ \mbox{erg/s}
\end{equation}
associated with a very low characteristic frequency,
\begin{equation}
\nu_{GW} \sim {\tau_{dyn}}^{-1}= 3 \times 10^{-17} \ \bigg (\frac{\bar{M}}{10^{10} \ M_\odot} \bigg )^{1/2} \ \bigg (\frac{\sigma_i}{25 \ \mbox{kpc}} \bigg )^{-3/2} \  \mbox{Hz}
\end{equation}
The luminosity $L_{GW}$ is comparable to the power radiated  by bright quasars in the electromagnetic  regime if $ \epsilon $ is of  order unity. 

Also, if the asymmetric kinetic energy $E_{nonsph}$ associated with a collapsing condensate at a distance $d$ from us is $ \sim G {\bar{M}}^2/\sigma_{min}$ then the ensuing GW amplitude on Earth can be estimated to be,
$$h_{GW} \sim \frac{4 G E_{nonsph}}{c^4 d}$$
\begin{equation}
 = \frac {2 }{d} \bigg (\frac {2 G \bar{M}} {c^2} \bigg ) \bigg (\frac {m \bar{M}}{m^2_{Pl}} \bigg )^2\sim 3 \times 10^{-10} \bigg (\frac{\bar{M}}{10^{10} \ M_\odot} \bigg )^{3} \ \bigg (\frac{m}{1.54 \times 10^{-20} \ \mbox{eV}} \bigg )^2 \bigg ( \frac{d}{10 \ \mbox{Mpc}}    \bigg )^{-1} \ \ \ .
\end{equation}
On the other hand, if the central BH is not formed, the dark bosons can form a stable self-gravitating system with a characteristic size $R_0=2 \sigma_{min} $  that minimizes $V_{eff}$ of eq.(53). Tidal forces, as  two galaxies go past each other,  can induce small oscillations in the condensate about the  size $R_0 $,
\begin{equation}
2 \sigma_{min}= 176 \  \bigg ( \frac {10^9 M_\odot}{M} \bigg ) \bigg ( \frac {10^{-22} \ \mbox{eV}}{m} \bigg )^2 \ \mbox{pc} 
\end{equation}
that follows from  eq.(56) (which essentially is not very different from the result of eq.(24) when $n=1$).

 The normal mode angular frequency $\omega_{osc} $ of such an undulation is given by,

\begin{equation}
\omega^2_{osc} =\frac {1}{m} \frac{d^2V_{eff}} {d\sigma^2}\bigg \vert_{R_0}=\frac { G^4 {\bar{M}}^4 m^6}{100 \ \hbar^6}
\end{equation}
corresponding to a frequency,
\begin{equation}
\nu_{osc}= \frac{\omega_{osc}} {2 \pi}= 1.35 \times 10^{-15} \ \bigg (\frac{\bar{M}}{10^{9} \ M_\odot} \bigg )^2 \ \bigg (\frac{m}{ 10^{-22} \ \mbox{eV}} \bigg )^3 \  \mbox{Hz} \ .
\end{equation}
Such oscillations could play an important role in amplifying star formation rates in the central region of interacting galaxies, as discussed in the preceding subsection. From eq.(71) it is evident that if  dark bosons have mass as low as $10^{-22}$ eV, the associated emission of gravitational  waves (GWs) will   presently be unmeasurable. This is because of the fact that the  current Pulsar Timing Arrays (PTA), which  use arrival  times of radio-pulses from milli-second pulsars, can   only detect GWs that have frequency $\geq$ nanoHertz \cite {pdg-ref-66}. 

However, in this scenario, DM particles with rest mass $ \geq 10^{-20}$ eV are promising not only in generating SMBHs with mass $\gtrsim 10^{10} M_\odot $ but also in entailing emission of GWs with frequency  $\gtrsim 10^{-8}  $ Hz from undulating BECs, whose existence can be constrained by the current PTA.   

\section*{Discussion}
 Existence of very light ($m \sim 10^{-23}- 10^{-20} \ \mbox{eV}$) bosonic DM particles can have far reaching astrophysical consequences - early formation of SMBHs with mass $> 10^{9} \ M_\odot$, enhancement of SFR in the nuclear region of interacting galaxies, GWs from dark BEC undergoing dynamical change   as well as dark energy.
 
Since BEC oscillation frequency is very sensitive to the rest mass of the dark boson,  detection of GWs from such undulations is possible in the near future only if $m \gtrsim 10^{-20}$ eV. While $m \sim 10^{-23}- 10^{-22}$ eV has interesting repercussions as far as star burst galaxies,  very high redshift host galaxies of AGNs with large amount of dust and DE are concerned, it leads to generation of SMBHs with mass $\gtrsim 10^{11} \ M_\odot$, in our model.

\section*{Acknowledgements}
It is a pleasure to thank Prof. Frank H. Shu and Dr. N. Rathnasree for their helpful comments. PDG also thanks Mr. Subhrendu Malakar and  Mr. Shrey Ansh for drawing his attention to two recent research papers on quasar J1342+0928.


\begin{thebibliography}{99}


\bibitem{pdg-ref-0}

Ho, L. C. (1999), {\it Observational Evidence for Black Holes in the Universe},  (Springer Netherlands) pp. 157-186

\bibitem{pdg-ref-1}

Ferrarese, L., \& Merritt, D. (2000), {\bf Ap J, 539}, L9


\bibitem{pdg-ref-2}

Kormendy, J., \& Ho, L. C. (2013), {\bf Ann. Rev. Astr. \& Ap, 51}, 511

\bibitem{pdg-ref-3}

Rees, M. J. (1984), {\bf Ann. Rev. Astr. \& Ap, 22}, 471

\bibitem{pdg-ref-4}

Wagh, S.M. and Dadhich, N. (1989) {\bf Phys. Rep., 183}, pp.137-192; and the references therein.

\bibitem{pdg-ref-5}

Das Gupta, P. (2015), {\it Astronomical Society of India Conference Series}, {\bf 12}, pp.1-8 

\bibitem{pdg-ref-6}

Barth A. J., Martini P., Nelson C. H.,  \& Ho L. C. (2003), {\bf Ap J, 594}, L95

\bibitem{pdg-ref-7}

Willott C. J., McLure R. J.,  \& Jarvis M. J. (2003), {\bf Ap J Lett, 587}, L15

\bibitem{pdg-ref-8}

Ghisellini G., et al. (2010), {\bf MNRAS, 405,} 387 

\bibitem{pdg-ref-9}

Wu, X. B. et al. (2015), {\bf Nature, 518}, 512

\bibitem{pdg-ref-10}

Venemans, Bram P. et al. (2017), {\bf Ap J Lett.,  851}, L8

\bibitem{pdg-ref-11}

Ba$\tilde {\mbox{n}}$ados et al. (2017), {\bf Nature}, http://dx.doi.org/10.1038/nature25180 (2017)

\bibitem{pdg-ref-12}

Bertone, G., Hooper, D., \& Silk, J. (2005), {\bf  Phys. Rep., 405}, 279

\bibitem{pdg-ref-13}

Spergel, D.N., Bean, R., Doré, O., Nolta, M.R., Bennett, C.L., Dunkley, J., Hinshaw, G., Jarosik, N.E., Komatsu, E., Page, L. \& Peiris, H.V. (2007),  {\bf Ap J Suppl. Ser., 170}, 377

\bibitem{pdg-ref-14}

Hui, L., Ostriker, J. P., Tremaine, S. \& Witten, E. (2017), {\bf Phys. Rev., D 95}, 043541; and the references therein. 

\bibitem{pdg-ref-15}

https://cds.cern.ch/record/2139643/files/ATLAS-CONF-2016-009.pdf

\bibitem{pdg-ref-16}



Tan, A., et al. (2016), {\bf Phys. Rev. Lett., 117}, 121303

\bibitem{pdg-ref-17}

Aprile, E., et al., 2016, {\bf Phys. Rev., D 94}, 122001

\bibitem{pdg-ref-18}

Akerib, D.S. et al. (2017), {\bf  Phys. Rev. Lett., 118}, 021303 

\bibitem{pdg-ref-19}

Schive, Hsi-Yu,   Chiueh, T  \&  Broadhurst, T. (2014), {\bf Nature Physics (cover), 10}, 496 

\bibitem{pdg-ref-20}

Das Gupta, P. (2009), arXiv:0905.1621

\bibitem{pdg-ref-21}

Das Gupta, P. \& Thareja, E. (2017), {\bf Class. Quan. Grav., 34}; and the references therein.

\bibitem{pdg-ref-22}

Madau, P., \& Rees, M. J. (2001), {\bf Ap J, 551}, L27

\bibitem{pdg-ref-23}

Volonteri, M., Haardt, F., \& Madau, P. (2003), {\bf Ap J, 582}, 559

\bibitem{pdg-ref-24}

Shapiro, S. L. (2004), {\bf Ap J, 610}, 913

\bibitem{pdg-ref-25}

Begelman, M. C., Volonteri, M., \& Rees, M. J. (2006), {\bf MNRAS, 370}, 289

\bibitem{pdg-ref-26}

Peirani, S., \& de Freitas Pacheco, J. A. (2008), {\bf Phys. Rev. D, 77}, 064023

\bibitem{pdg-ref-27}

Natarajan P.,  \& Treister E. (2009), {\bf MNRAS, 393}, 838

\bibitem{pdg-ref-28}

Mayer, L., Kazantzidis, S., Escala, A., \& Callegari, S. (2010), {\bf Nature, 466}, 1082

\bibitem{pdg-ref-29}

Johnson, J. L. (2013), {\it First Galaxies: Theoretical Predictions and Observational Clues}, ed. T. Wiklind, B. Mobasher \& V.  Bromm (Springer-Verlag Berlin Heidelberg),  177

\bibitem{pdg-ref-30}

Walker S. A., Fabian A. C., Russell H. R.,  \& Sanders J. S. (2014), {\bf MNRAS, 442}, 2809

\bibitem{pdg-ref-31}

King, A. (2016), {\bf MNRAS, 456}, L109




\bibitem{pdg-ref-32}

Abbott, B.P., Abbott, R., Abbott, T.D., Abernathy, M.R., Acernese, F., Ackley, K., Adams, C., Adams, T., Addesso, P., Adhikari, R.X. \& Adya, V.B. (2016), {\bf Phys. Rev. Lett., 116}, 241102


\bibitem{pdg-ref-33}

Abbott, B.P., Abbott, R., Abbott, T.D., Abernathy, M.R., Acernese, F., Ackley, K., Adams, C., Adams, T., Addesso, P., Adhikari, R.X. \& Adya, V.B. (2016), {\bf Phys. Rev. Lett., 116}, 241103  

\bibitem{pdg-ref-34}

Abbott, B.P., Abbott, R., Abbott, T.D., Acernese, F., Ackley, K., Adams, C., Adams, T., Addesso, P., Adhikari, R.X., Adya, V.B. \& Affeldt, C. (2017), arXiv:1706.01812

\bibitem{pdg-ref-35}

Abbott, B.P., Abbott, R., Abbott, T.D., Acernese, F., Ackley, K., Adams, C., Adams, T., Addesso, P., Adhikari, R.X., Adya, V.B. and Affeldt, C. (2017), {\bf Phys. Rev. Lett., 119}, 141101 

\bibitem{pdg-ref-36}

Bardeen, J.M., Carter, B. \& Hawking, S.W. (1973), {\bf Comm. Math. Phys., 31}, 161

\bibitem{pdg-ref-37}

Das Gupta, P., (2016) {\bf Journal of Scientific Review, 6}(1)

\bibitem{pdg-ref-38}

Thorne,  K. S.  (1983), in Gravitational Radiation, eds. N. Deruelle and T. Piran
(Amsterdam: North-Holland)

\bibitem{pdg-ref-39}

Dyson, F. (1963), in Interstellar Communication, ed. A.G. Cameron (New
York: Benjamin)


\bibitem{pdg-ref-40}

Hogan, C. J. (2001), in Structure Formation in the Universe (Springer Netherlands)

\bibitem{pdg-ref-41}

Barrow, J. D. \&  Gibbons G. W. (2017), {\bf Phys. Rev.D, 95},  064040 

\bibitem{pdg-ref-42}

Hawking, S.W., Moss, I.G. \& Stewart, J.M. (1982)  {\bf Phys. Rev. D, 26}, 2681

\bibitem{pdg-ref-43}

Upadhyay, N.,  Das Gupta, P. \&  Saxena, R. P. (1999) {\bf Phys. Rev. D, 60}, 063513; and the references therein.

\bibitem{pdg-ref-44}

Das Gupta, P. \& Saini, N., to appear in the {\bf Journal of Astrophysics and Astronomy}, arXiv:1709.00185v1 [astro-ph.HE] 



\bibitem{pdg-ref-45}

Abbott, B.P., Abbott, R., Abbott, T.D., Abernathy, M.R., Acernese, F., Ackley, K., Adams, C., Adams, T., Addesso, P., Adhikari, R.X. \& Adya, V.B. (2017), {\bf Annalen der Physik, 529}, 1

\bibitem{pdg-ref-46}

Hawking, S. W. (1974), {\bf Nature, 248}, 30 

\bibitem{pdg-ref-47}

Das Gupta, P. (2010), {\bf Radiation Effects \& Defects in Solids, 165}, pp.106-113


\bibitem{pdg-ref-48}


Fukuyama, T. \& Morikawa, M. (2007), in {\it Relativistic Astrophysics and Cosmology - Einstein's Legacy}, eds. B. Aschenbach, V. Burwitz, G. Hasinger \& B. Leibundgut (Springer Berlin Heidelberg), 95


\bibitem{pdg-ref-49}

Das, S., \& Bhaduri, R. K. (2015), {\bf Class. Quan. Grav., 32}, 105003


\bibitem{pdg-ref-50}

Hoyle, F. (1949), in {\it Problems of Cosmical Aerodynamics}, ed. J. M. Burgers \& H. C. van de Hulst (International Union of Theoretical and Applied Mechanics, and International Astronomical Union), 195

\bibitem{pdg-ref-51}

Peebles, P.J.E. (1969),  {\bf Ap J, 155}, 393


\bibitem{pdg-ref-52}

Larson, R.B. \& Tinsley, B.M. (1978),  {\bf Ap J, 219}, 46

\bibitem{pdg-ref-53}

Mihos, C. \& Hernquist, L. (1996), {\bf Ap J, 464}, 641; and the references therein. 

\bibitem{pdg-ref-54}

Kewley, L.J., Dopita, M.A., Sutherland, R.S., Heisler, C.A. \& Trevena, J. (2001), {\bf Ap J, 556}, 121; and the references therein.


\bibitem{pdg-ref-55}

Groves, B., Dopita, M.A., Sutherland, R.S., Kewley, L.J., Fischera, J., Leitherer, C., Brandl, B. \& van Breugel, W. (2008), {\bf Ap J Supp. Ser., 176}, 438; and the references therein.



\bibitem{pdg-ref-56}

Lynden-Bell, D. (1967), {\bf MNRAS, 136}, 101

\bibitem{pdg-ref-57}

Arad, I. \& Lynden-Bell, D. (2005), {\bf MNRAS, 361}, 385

\bibitem{pdg-ref-58}

Byrd, G. \& Valtonen, M. (1990), {\bf Ap J, 350}, 89


\bibitem{pdg-ref-59}

Shu, F.H., Milione, V., Gebel, W., Yuan, C., Goldsmith, D.W. \& Roberts, W.W. (1972), {\bf Ap J, 173}, 557

\bibitem{pdg-ref-60}

Hu, W.,  Barkana, R. \&  Gruzinov, A. (2000), {\bf Phys. Rev. Lett., 85}, 1158


\bibitem{pdg-ref-61}

Perez-Garcia, V. M., Michinel, H., Cirac,  J. I., Lewenstein, M. \& Zoller, P. (1996), {\bf Phys. Rev. Lett., 77}, 5320

\bibitem{pdg-ref-62}

Perez-Garcia, V. M., Michinel, H., Cirac,  J. I., Lewenstein, M. \& Zoller, P.  (1997),  {\bf Phys. Rev. A, 56}, 1424


\bibitem{pdg-ref-63}

Das Gupta, P. (2015), {\bf Current Science, 109}, 1946


\bibitem{pdg-ref-64}

Das Gupta, P. \& Rahman, F, in preparation.




\bibitem{pdg-ref-65}

Bogdan, A. \& Goulding, A.D. (2015), {\bf Ap J, 800}, 124


\bibitem{pdg-ref-66}


Mingarelli, C.M., Lazio, T.J.W., Sesana, A., Greene, J.E., Ellis, J.A., Ma, C.P., Croft, S., Burke-Spolaor, S. \& Taylor, S.R. (2017), {\bf Nature Astronomy, 1}, 886; and the references therein.


\end{thebibliography}
\end{document}